\documentclass[conference]{IEEEtran}
\IEEEoverridecommandlockouts

\usepackage{cite}
\usepackage{amsmath,amssymb,amsfonts}
\usepackage{algorithmic}
\usepackage{graphicx}
\usepackage{textcomp}
\usepackage{xcolor}
\usepackage{multirow}
\usepackage{jabbrv}

\def\BibTeX{{\rm B\kern-.05em{\sc i\kern-.025em b}\kern-.08em
    T\kern-.1667em\lower.7ex\hbox{E}\kern-.125emX}}

\makeatletter 
\newcommand{\linebreakand}{%
\end{@IEEEauthorhalign}
\hfill\mbox{}\par
\mbox{}\hfill\begin{@IEEEauthorhalign}
}
\makeatother 

\begin{document}


\title{Towards Dynamic Neural Communication and Speech Neuroprosthesis Based on Viseme Decoding\\
}

\author{\IEEEauthorblockN{Ji-Ha Park}
\IEEEauthorblockA{\textit{Dept. of Artificial Intelligence} \\
\textit{Korea University} \\
Seoul, Republic of Korea \\
jiha\_park@korea.ac.kr}
\\
\IEEEauthorblockN{Soowon Kim}
\IEEEauthorblockA{\textit{Dept. of Artificial Intelligence} \\
\textit{Korea University} \\
Seoul, Republic of Korea \\
soowon\_kim@korea.ac.kr}
\and
\IEEEauthorblockN{Seo-Hyun Lee}
\IEEEauthorblockA{\textit{Dept. of Brain and Cognitive Engineering} \\
\textit{Korea University} \\
Seoul, Republic of Korea \\
seohyunlee@korea.ac.kr}
\\
\IEEEauthorblockN{Seong-Whan Lee}
\IEEEauthorblockA{\textit{Dept. of Artificial Intelligence} \\
\textit{Korea University} \\
Seoul, Republic of Korea \\
sw.lee@korea.ac.kr}
}

\maketitle




\begin{abstract}
Decoding text, speech, or images from human neural signals holds promising potential both as neuroprosthesis for patients and as innovative communication tools for general users. Although neural signals contain various information on speech intentions, movements, and phonetic details, generating informative outputs from them remains challenging, with mostly focusing on decoding short intentions or producing fragmented outputs. In this study, we developed a diffusion model-based framework to decode visual speech intentions from speech-related non-invasive brain signals, to facilitate face-to-face neural communication. We designed an experiment to consolidate various phonemes to train visemes of each phoneme, aiming to learn the representation of corresponding lip formations from neural signals. By decoding visemes from both isolated trials and continuous sentences, we successfully reconstructed coherent lip movements, effectively bridging the gap between brain signals and dynamic visual interfaces. The results highlight the potential of viseme decoding and talking face reconstruction from human neural signals, marking a significant step toward dynamic neural communication systems and speech neuroprosthesis for patients.
\end{abstract}




\section{INTRODUCTION}

Brain-computer interface (BCI) is a technology that interprets and conveys the human mind by translating brain signals that encapsulate various aspects of user intention and cognition~\cite{ding2013changes, karikari2023review, jeong2019classification, peksa2023state, won2020adaptive}. Recent advances in signal processing and generative technologies are being integrated with BCI, giving rise to a novel form of intuitive neural communication represented by speech BCI~\cite{chaudhary2016brain}. This technology aims to decode speech-related neural signals and directly generate text or speech~\cite{lee2020neural,lee2022toward}. Speech BCIs are an important and rapidly emerging field with significant potential, not only as assistive technologies for individuals with speech impairments, such as those with aphasia but also as a groundbreaking mode of neural communication for general users~\cite{moses2021neuroprosthesis, willett2023high, anumanchipalli2019speech, karikari2023review}.

\begin{figure*}[t]
\centerline{\includegraphics[width=\textwidth]{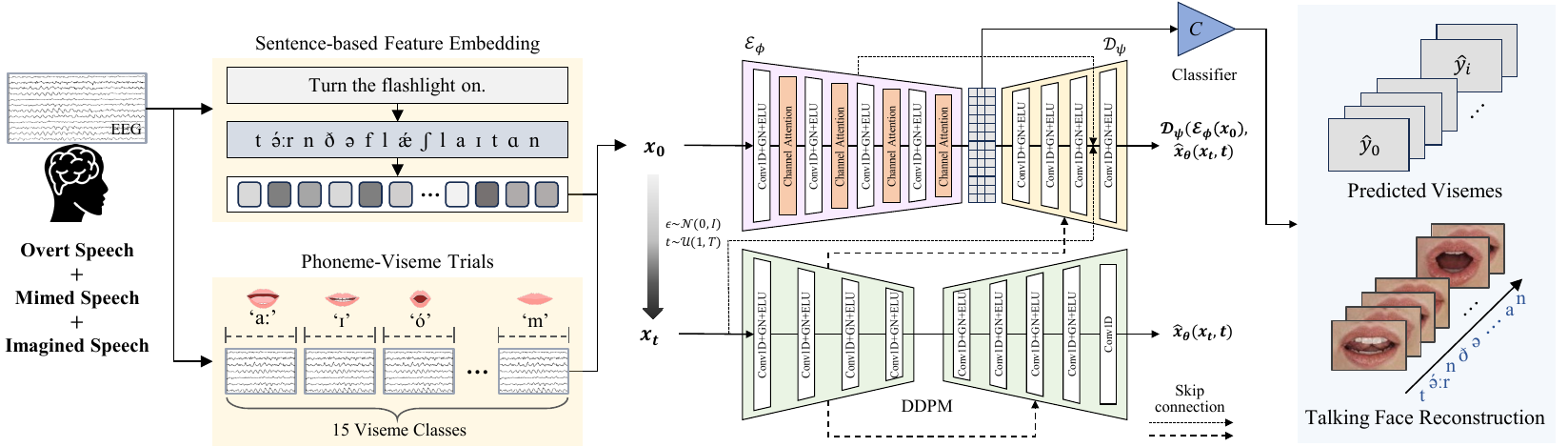}}
\caption{Overview of the proposed viseme decoding framework from EEG signals. The recorded EEG signals are preprocessed in two ways for training. Firstly, sentence-level EEG signals are segmented at the phoneme-level based on real audio of uttered sentences to extract phoneme-based viseme feature embeddings. Secondly, single-trial EEG data of viseme classes are preprocessed and given as input features. DDPM in the proposed framework learns relevant features by injecting noise and restoring the data, while the encoder and decoder work together to compensate for the information loss during this process, allowing the accurate reconstruction of EEG signals. The trained model then predicts the viseme classes by passing the compressed information from the final encoder stage through a classifier. Finally, the predicted viseme classes are sequentially reconstructed into articulatory movements of sentences.}
\label{fig1}
\end{figure*}

At the same time, advancements in computer vision and speech processing methods have significantly enhanced the generation of realistic talking faces that synchronize with users' speech and facial expressions, providing immersive and interactive visual experiences~\cite{prajwal2020lip, ji2021audio, zhou2018visemenet, eldawlatly2024role, vougioukas2020realistic}. The convergence of these technologies with BCI may open new avenues of exploring how visual speech intentions, such as facial movements and expressions, can be decoded directly from brain signals and transformed into dynamic visual outputs~\cite{park2024towards, takagi2023high, metzger2023high}. This opens up the possibility of expressive neural communication for general users, and significant speech neuroprostheses for language-impaired patients that can provide speech-visual feedback to facilitate rehabilitation~\cite{willett2023high, moses2021neuroprosthesis}.


Despite these promising potentials, most studies have focused on generating fragmented or abstract outputs from neural signals, as challenges remain in translating these signals into dynamic and unconstrained forms~\cite{lee2019towards}. Especially, extracting relevant features to reconstruct complex lip movements or subtle facial expressions from noisy neural signals poses a significant challenge with great potential for advancement. While signal processing techniques using invasive measurements have demonstrated their potential for robust decoding of speech intentions, their reliance on surgical procedures limits their application~\cite{willett2023high, metzger2023high}. Non-invasive methods such as electroencephalography (EEG) remain underexplored due to their low signal-to-noise ratio (SNR), despite offering significantly broader applicability and potential benefits~\cite{mane2021fbcnet,menon2005combined, lee2023AAAI}. Therefore, developing techniques that can effectively process non-invasive neural signals despite low SNR is crucial for adaptable neural communication and speech neuroprosthesis~\cite{park2024towards}. To achieve this, a novel approach capable of robustly capturing subtle information is necessary to enhance both the fidelity and the ability to generate unconstrained visual representations from non-invasive brain signals.


\begin{figure}[t]
\centerline{\includegraphics[width=\columnwidth]{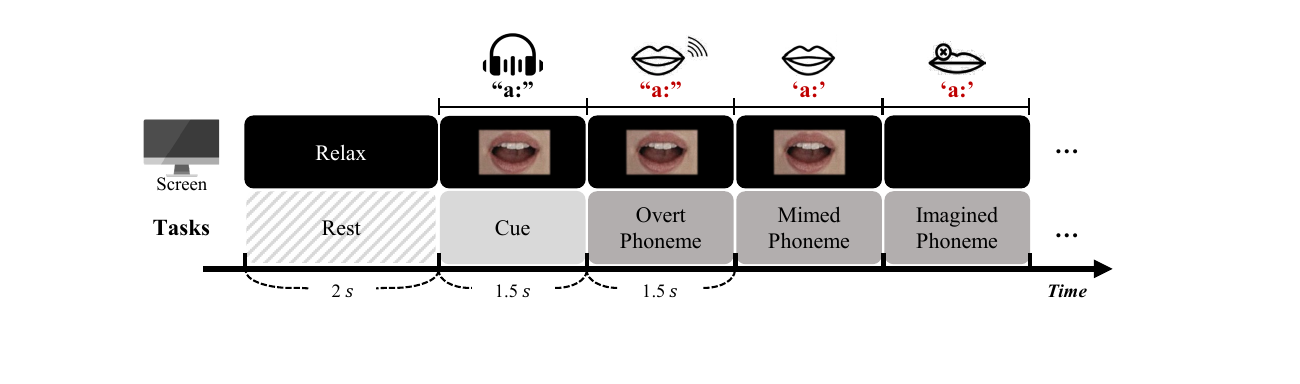}}
\caption{Experimental paradigm for EEG signals recording. The phonemes and sentences were presented randomly to guide overt, mimed, and imagined speech. Participants were instructed to perform the experiment with a focus on the changes in lip shapes and articulator movements.}
\label{fig2}
\end{figure}

In this paper, we introduce a diffusion model-based viseme decoding framework that effectively learns visual speech intentions from speech-related EEG signals. The viseme-level decoding demonstrates the potential to generate dynamic outputs from noisy and limited non-invasive neural signals, advancing beyond conventional word- or sentence-level decoding, and offering potential applications for patients with progressive paralysis. Our generative model-based framework goes beyond predefined class decoding, enabling robust decoding through phoneme-level clustering while extending to unconstrained reconstruction by learning small fragments of speech.

\section{METHODS}

\subsection{Model Architectures} 
\subsubsection{Denoising Diffusion Probabilistic Models}
The overall structure of the model consists of denoising diffusion probabilistic models (DDPMs), a conditional autoencoder (CAE), and a classifier (Fig.~\ref{fig1})~\cite{kim2023diff}. DDPMs are based on a time-conditional U-Net architecture, designed to learn how to represent EEG data by progressively adding noise in a stepwise manner~\cite{ho2020denoising}. Unlike the conventional technique of training a model with a noisy $\mathbf{x}_t$ and predicting the noise it contains by training network $\epsilon\theta(\mathbf{x}_t,t)$, our model is trained to reach the original signal $\mathbf{x}_0$:
\begin{equation}
\mathcal{L}_{\text{DDPM}}(\theta) = ||\mathbf{x}_0 - \hat{\mathbf{x}}_\theta(\mathbf{x}_t,t)||
\label{eq:DDPM}
\end{equation}

\subsubsection{Conditional Autoencoder}
The information loss of the DDPM is identified and corrected by the CAE~\cite{zhang2022unsupervised}. The CAE is composed of an encoder and decoder, denoted as $\mathcal{E}_\phi$ and $\mathcal{D}_\psi$, respectively (Fig.~\ref{fig1}). Instead of connecting to the output of $\mathcal{E}_\phi$, $\mathcal{D}_\psi$ is skip-connected with the DDPM layers, enabling $\mathcal{D}_\psi$ to be indirectly influenced by the corruption stage of the DDPM. Additionally, to improve the reconstruction of $\mathcal{L}_{\text{DDPM}}$, the original signal $\mathbf{x}_0$ and the DDPM output $\hat{\mathbf{x}}_\theta(\mathbf{x}_t,t)$ are skip-connected to the final layer of $\mathcal{D}_\psi$. To ensure that $\mathcal{E}_\phi$ effectively derives meaningful representations related to visemes, channel attentions are applied after each layer block of $\mathcal{E}_\phi$. This structure learns the importance of each channel in the input features, calculates channel weights, and multiplies them with the original input to generate a weighted output. By doing so, the CAE can generate more accurate representations of the original signals.
\begin{equation}
    \mathcal{L}_{\text{CAE}}(\psi, \phi) = ||\mathcal{L}_{\text{DDPM}}(\theta) -\mathcal{D}_\psi(\mathcal{E}_\phi(\mathbf{x}_0),\hat{\mathbf{x}}_\theta(\mathbf{x}_t, t))||
\label{eq:CAE}
\end{equation}

\subsubsection{Viseme Decoding Process}
The output from $\mathcal{E}_\phi$ is compressed into a one-dimensional representation $\mathbf{z}$ through an adaptive average pooling layer. This latent vector is then passed to the linear classifier $\mathcal{C}_\rho$. $\mathcal{C}_\rho$, is composed of Kolmogorov–Arnold Networks (KAN), which incorporate linear layers and group normalization~\cite{liu2024kan}.
$\mathcal{C}_\rho$ is co-trained with the CAE to distinguish between class representations and perform classification. Only $\mathcal{E}_\phi$ and $\mathcal{C}_\rho$ are employed to classify the signals. Finally, the objective function of the CAE is integrated and the total classification loss is defined as follows:
\begin{equation}
    \begin{split}
        \mathcal{L}_{\text{total}}(\psi, \phi, \rho) &= \mathcal{L}_{\text{CAE}}(\psi, \phi) +\alpha||\hat{y}-y||_2
    \end{split}
\label{eq:total_cls}
\end{equation}


\begin{table}[t]
\setlength{\tabcolsep}{2.4pt}
\renewcommand{\arraystretch}{1}
\caption{The average viseme decoding performance across all subjects in overt, mimed, and imagined speech.}
\centering
\begin{tabular}{ccccc}
\hline
                                   &                & \textbf{VER (\%) $\downarrow$}   & \textbf{F1-score $\uparrow$}    & \textbf{AUC (\%) $\uparrow$}   \\ \hline
\multirow{4}{*}{\textbf{Overt}}    & EEGNet         & 44.82 $\pm$ 10.60         & 54.53 $\pm$ 10.82          & 89.15 $\pm$ 4.59          \\
                                   & DeepConvNet    & 52.96 $\pm$ 10.72         & 45.94 $\pm$ 11.02          & 86.82 $\pm$ 4.74          \\
                                   & ShallowConvNet & 41.85 $\pm$ 9.95          & 55.18 $\pm$ 9.59           & 91.83 $\pm$ 4.03          \\
                                   & \textbf{Ours}  & \textbf{34.07 $\pm$ 5.34} & \textbf{65.85 $\pm$ 5.63}  & \textbf{93.55 $\pm$ 2.16} \\ \hline
\multirow{4}{*}{\textbf{Mimed}}    & EEGNet         & 61.11 $\pm$ 6.26          & 37.10 $\pm$ 6.20           & 82.56 $\pm$ 6.18          \\
                                   & DeepConvNet    & 65.37 $\pm$ 8.83          & 33.41 $\pm$ 9.32           & 78.29 $\pm$ 6.57          \\
                                   & ShallowConvNet & 55.74 $\pm$ 6.63          & 41.42 $\pm$ 7.65           & 84.96 $\pm$ 3.71          \\
                                   & \textbf{Ours}  & \textbf{48.33 $\pm$ 6.96} & \textbf{51.54 $\pm$ 6.52}  & \textbf{87.38 $\pm$ 5.23} \\ \hline
\multirow{4}{*}{\textbf{Imagined}} & EEGNet         & 83.71 $\pm$ 3.21          & 14.54 $\pm$ 1.49           & 58.32 $\pm$ 7.42          \\
                                   & DeepConvNet    & 83.15 $\pm$ 6.10          & 15.89 $\pm$ 6.10           & 57.50 $\pm$ 7.66          \\
                                   & ShallowConvNet & 82.22 $\pm$ 3.65          & 14.04 $\pm$ 4.76           & 61.05 $\pm$ 8.71          \\
                                   & \textbf{Ours}  & \textbf{77.96 $\pm$ 9.78} & \textbf{21.04 $\pm$ 10.25} & \textbf{66.67 $\pm$ 9.55} \\ \hline
\end{tabular}
\label{tab1}
\end{table}

\subsubsection{Sentence-level Reconstruction}
The model was fine-tuned to capture the continuous viseme sequences from the sentence-level EEG signals. The model, initially optimized on isolated trials, is subsequently trained with EEG signal segments that are epoched based on phoneme units derived from the recorded audio signals of spoken sentences~\cite{mcauliffe2017montreal}. During this process, the weights of the final down-sampling layer of the $\mathcal{E}_\phi$ are updated, while other components remain frozen. This preserves the reconstruction capability of the pre-trained DDPM and $\mathcal{D}_\psi$ for EEG signals while allowing the fine-tuned $\mathcal{E}_\phi$ to extract more salient and meaningful features.

\subsection{Experimental Details} 

\subsubsection{Data Recordings}
EEG signals were recorded using a high-density EEG cap with active Ag/AgCl electrodes arranged in 128 channels at a sampling rate of 1,000~Hz. Fifteen distinctive viseme classes were defined, corresponding to 39 phonemes~\cite{pandzic2003mpeg}. The dataset comprised 8,100 isolated trials and 7,629 trials from sentence-level windowing from three subjects, each isolated trials lasting 1.5 seconds for each viseme class (Fig.~\ref{fig2}).
To leverage sentence-level EEG data for viseme training, ten different sentences were recorded, with each sentence lasting 3 seconds and repeated 5 times across the trials.
Each subject participated in a series of overt, mimed, and imagined speech. We collected EEG data along with corresponding audio and video, capturing the neural, auditory, and visual components related to visemes. The experiment was conducted in accordance with the Declaration of Helsinki, approved by the Korea University Institutional Review Board [KUIRB-2022-0104-03].


\begin{figure}[t]
\centerline{\includegraphics[width=0.915\columnwidth]{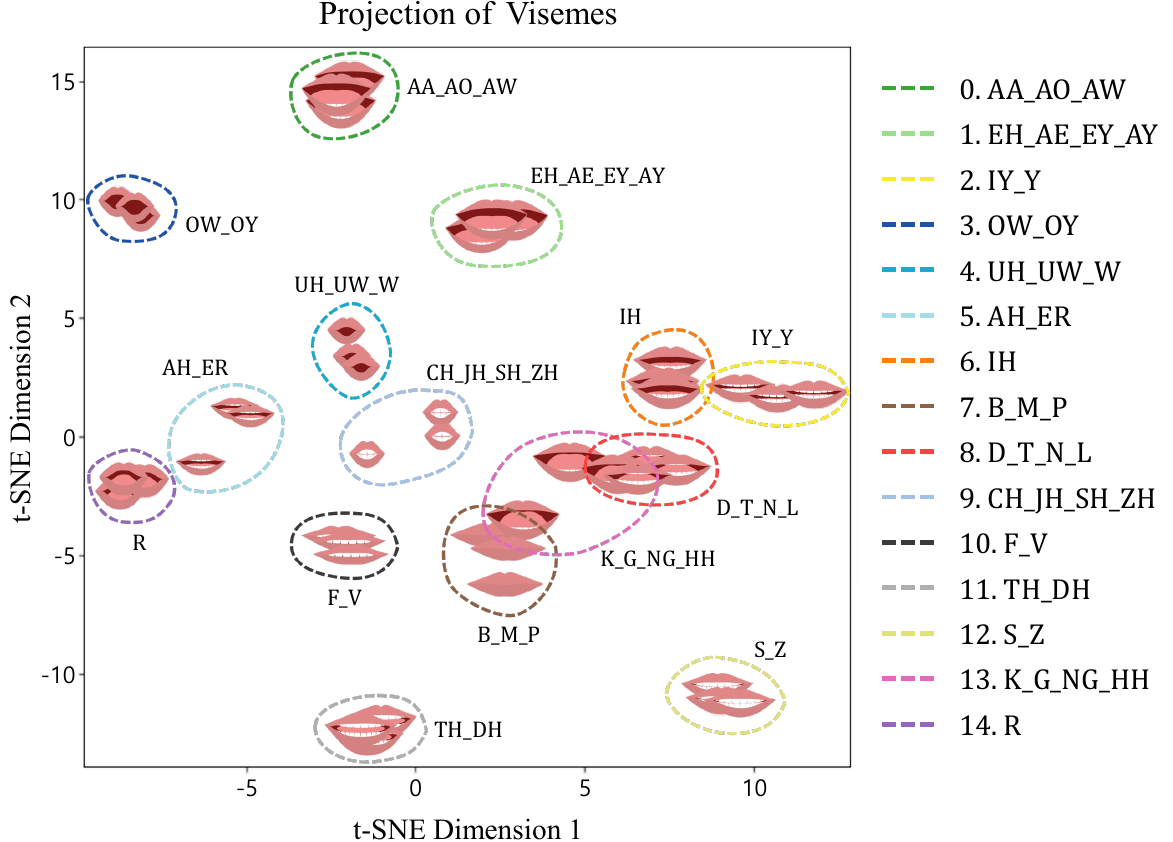}}
\caption{The projection of phoneme encodings into a 2D space using t-SNE shows that phoneme classes representing visemes clustered together, with class groups exhibiting similar lip shapes positioned relatively close to each other.
}
\label{fig3}
\end{figure}

\subsubsection{Signal Preprocessing and Training Details}
We applied a 5th-order Butterworth bandpass filter in the range of 30--499 Hz containing speech-related information in brain signals~\cite{lee2020neural,suk2014predicting}. Additionally, notch filtering was employed at harmonics of 60~Hz to remove the line noise. Signal preprocessing was conducted in Python and Matlab, using the OpenBMI Toolbox~\cite{leeMH2019eeg}, BBCI Toolbox~\cite{krepki2007berlin}, and EEGLAB Toolbox~\cite{delorme2004eeglab}. To overcome the limitations of training visemes on restricted datasets and short fragments, we leveraged sentence-level EEG data for viseme training. This approach enabled the capture of visemes from continuous EEG at the sentence level, providing a more robust method for viseme decoding. The dataset was randomly split into 8:2 in training and test sets and trained with a fixed random seed for consistency. The backbone architecture training details followed Kim et al.~\cite{kim2023diff}.

\begin{figure*}[t]
\centerline{\includegraphics[width=\textwidth]{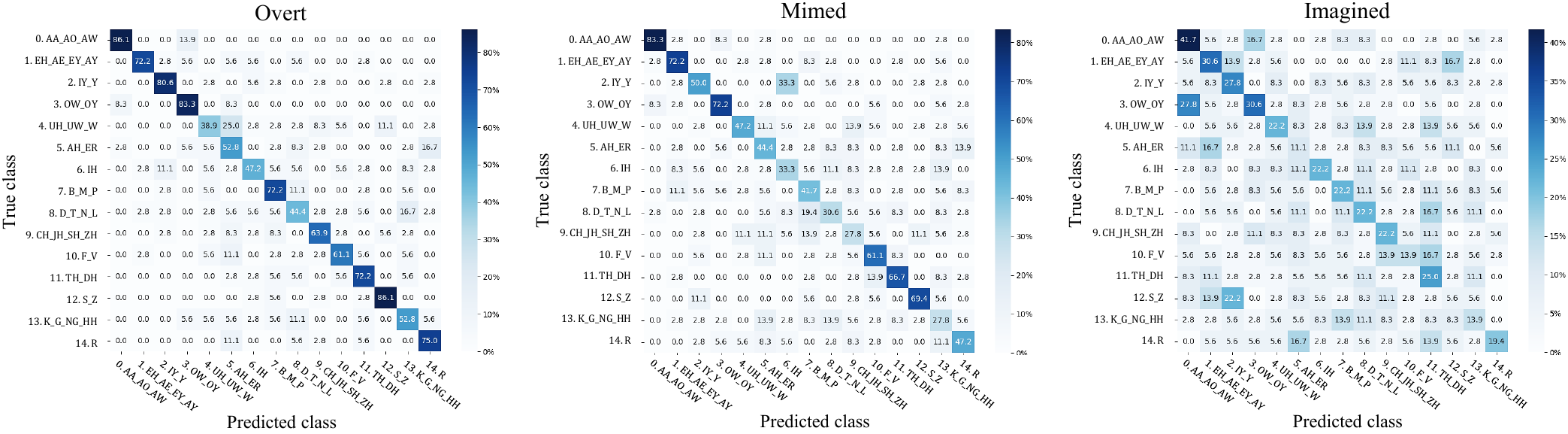}}
\caption{The confusion matrices of the classification results across overt, mimed, and imagined speech. The results indicated that overt speech displays more distinguishable patterns. Even in mimed and imagined speech, where there was no audible speech and reduced articulatory cues, our approach could capture and differentiate subtle patterns, supporting the proposed framework.}
\label{fig4}
\end{figure*}

\begin{figure}[t]
\centerline{\includegraphics[width=0.97\columnwidth]{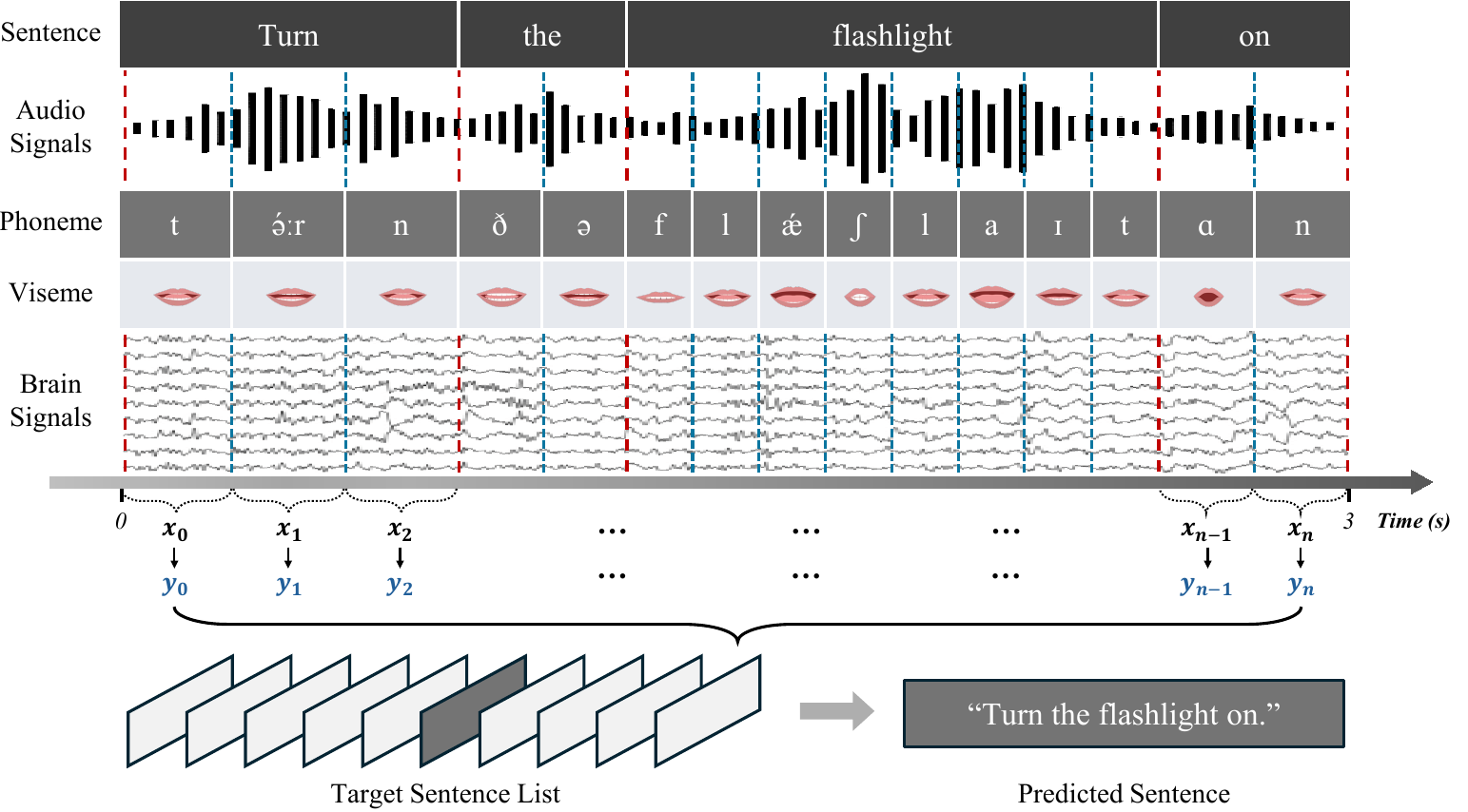}}
\caption{The recorded sentence data were epoched into phoneme-viseme segments based on the audio, which were used to fine-tune and adapt to the single trial-based trained model. The fine-tuned model, sharing similar articulatory and lip movement features, provided improved viseme decoding performance from continuous sentence-level brain signals. This allowed for more accurate reaching of the target sentence.}
\label{fig5}
\end{figure}

\section{RESULTS AND DISCUSSION}

\subsection{Viseme Decoding}
Table~\ref{tab1} displays the average decoding performance across all subjects in overt, mimed, and imagined speech. The evaluation metrics include the viseme error rate (VER), F1-score, and area under the curve (AUC).
Overt speech achieved the lowest VER of 34.07\%, showcasing strong viseme decoding, while mimed speech, solely relying on lip shapes without auditory output, reached 48.33\%. For imagined speech, without external articulation, a VER of 77.96\% highlighted the potential to decode internal speech from brain signals.
Our method consistently outperformed established networks, such as EEGNet~\cite{lawhern2018eegnet}, DeepConvNet~\cite{schirrmeister2017deep}, and ShallowConvNet~\cite{schirrmeister2017deep}, which have been reported to show robust performance in EEG signal decoding. The ability to decode fragmented segments of EEG signals into small viseme units suggests potential for unconstrained communication~\cite{gaddy2020digital}. Also, the ability to decode visemes without overt articulation underscores the system’s capability to capture and interpret internal cognitive processes related to speech tasks, paving the way for further advancements in silent communication~\cite{panachakel2021decoding, proix2022imagined}.

\subsection{Phoneme-Viseme Clustering}
Fig.~\ref{fig3} presents the t-SNE projection of phoneme encodings from EEG signals into a 2D space, revealing clear clustering patterns based on the corresponding viseme classes. The 39 phonemes were grouped together by their shared lip shapes when articulated, supporting the coherence of these phonetic groupings. Notably, viseme classes with subtle differences were positioned relatively close together, reflecting their phonetic similarity, whereas classes with more pronounced differences in articulation, such as rounded versus unrounded lip shapes, were clearly separated with minimal overlap. These results provide qualitative validation for the classification of 15 distinct viseme categories, indicating that the reduced set of viseme classes adequately captures the variability in lip shapes associated with different phonemes.
Furthermore, the clustering patterns underscore that EEG signals effectively encode speech-related lip shape information, demonstrating the model’s capacity to distinguish between varying viseme classes based on neural activity.
Fig.~\ref{fig4} shows the confusion matrices for each speech task, illustrating the classification performance across different viseme classes. The balanced classification performance across all classes further implied that our study consistently reads user visual speech intentions through viseme decoding across varying levels of speech, providing a foundation for future development in EEG-based visual speech interfaces~\cite{chartier2018encoding}.





\subsection{Face-to-face Neural Communication}
We extended our approach to continuous sentence-level data, moving beyond isolated trial classification (Fig.~\ref{fig5}). In natural sentence articulation, lip movements change rapidly, therefore, we inferred these changes in short time-step segments, sequentially combining the classification outputs for each segment and reconstructing continuous viseme sequences. By aligning the decoded visemes with the timing and context of actual sentence articulation, our system successfully produced outputs that were not only visually consistent but also highly realistic in terms of lip synchronization and expressiveness. This demonstrated that the proposed framework holds the potential to decode and reconstruct speech-related movements from continuous brain signals, even for previously unseen sentences~\cite{anumanchipalli2019speech}. For the predefined sentences used in the experiment, the system successfully inferred them in their complete form, proving that multi-output generation in text or audio formats could be feasible~\cite{lee2023AAAI}. While further research is needed, these findings provide evidence for the feasibility of face-to-face neural communication, demonstrating that our approach can serve as a visual interface capable of accurately capturing and expressing users' visual speech intentions directly from brain signals~\cite{metzger2023high}.

\section{CONCLUSION}
We proposed a diffusion-based viseme decoding framework that learns visual speech intentions from non-invasive speech-related brain signals. By mapping phonemes into viseme units, our model could capture subtle lip movements and articulation from noisy EEG signals. This allowed for realistic sentence-level reconstruction, generating continuous visemes. The viseme-level decoding showed potential for dynamic and unconstrained output generation from limited neural signals, moving beyond conventional word or sentence decoding. Our study advances the potential of neural signal-based communication systems, particularly for face-to-face interaction using speech-related non-invasive brain signals. The findings may mark a step toward developing intuitive neural communication systems and speech neuroprostheses.




\section*{Acknowledgment}

This work was partly supported by Institute of Information \& Communications Technology Planning \& Evaluation (IITP) grant funded by the Korea government (MSIT) (No. RS-2019-II190079, Artificial Intelligence Graduate School Program (Korea University), No. RS-2021-II-212068, Artificial Intelligence Innovation Hub, and No. RS-2024-00336673, AI Technology for Interactive Communication of Language Impaired Individuals).

\bibliographystyle{jabbrv_IEEEtran}
\bibliography{REFERENCE_jh}

\end{document}